\documentclass[fleqn,usenatbib]{mnras}

\usepackage{newtxtext,newtxmath}
\usepackage[T1]{fontenc}

\DeclareRobustCommand{\VAN}[3]{#2}
\let\VANthebibliography\thebibliography
\def\thebibliography{\DeclareRobustCommand{\VAN}[3]{##3}\VANthebibliography}

\usepackage{titlesec}
\titleformat*{\section}{\normalsize\bfseries}
\titleformat*{\subsection}{\normalsize\bfseries}
\usepackage{amsmath}
\usepackage{graphicx}
\usepackage{floatrow}
\usepackage{wrapfig}
\usepackage{xcolor}
\usepackage[bottom]{footmisc}
\usepackage{makecell}

\usepackage{graphicx}	% Including figure files
\usepackage{amsmath}	% Advanced maths commands	% Extra maths symbols
\usepackage{subfig}
\usepackage{newtxtext,newtxmath}
\title[The Optical Depth of 21 cm Foregrounds]{The Optical Depth of Foregrounds for the Highest Redshift 21 cm Signals}

\author[D. Seitova et al.]{
Daniya Seitova,$^{1}$\thanks{E-mail: daniya\_seitova@brown.edu}
Jonathan C. Pober,$^{1}$
\\
$^{1}$Department of Physics, Brown University, Providence RI 02912 USA 
}

\date{Accepted 2022 April 29. Received 2022 April 09; in original form 2021 November 17}

\pubyear{2022}

\begin{document}
\label{firstpage}
\pagerange{\pageref{firstpage}--\pageref{lastpage}}
\maketitle

\begin{abstract}
Foreground emission makes it difficult to detect the highly-redshifted cosmological 21\,cm signal at any frequency. However, at low frequencies foregrounds are likely to become optically thick, which would make it completely impossible to see a 21\,cm signal behind them. To find out which regions of the sky might be optically thick for the highest redshifts of the 21\,cm signal, we fit the measurements from LWA1 and the Haslam 408\,MHz map with  a two-component spectral model and calculate the frequency-dependent foreground optical depth point-by-point across the sky. Limitations of the current data prevent us from making any strong conclusions at high statistical significance, but there is suggestive evidence ($\sim1\sigma$) that as much as 25\% of the sky could be obscured for the highest redshift 21\,cm signals.
\end{abstract}


%%%%%%%%%%%%%%%%%%%%%%%%%%%%%%%%%%%%%%%%%%%%%%%%%%

%%%%%%%%%%%%%%%%% BODY OF PAPER %%%%%%%%%%%%%%%%%%

\newcommand{\fixfootnotes}{%
\renewcommand{\thefootnote}{\arabic{footnote}}%
\setcounter{footnote}{0}%
}
\fixfootnotess

\section{Introduction}

One of the most poorly constrained periods in the history of the universe is the time between the release of the Cosmic Microwave Background (CMB) radiation and the Epoch of Reionization, known as the Cosmic Dark Ages. No luminous sources existed during this epoch, making it exceedingly challenging to observe. However, the Dark Ages is an important transition period in cosmic history: the universe evolved from hydrogen and helium atoms to the large scale structure we see today. Studying the Dark Ages would provide a unique view of the physics of the early universe, when perturbations in the density field remain linear and can be modeled with well-understood physics. The only potential observational probe of this period, however, is through hydrogen’s 21\,cm line. 
In the standard cosmological picture, the hydrogen gas (and its ``spin temperature'', which characterizes the relative populations of its hyperfine ground state) is colder than the CMB during the Dark Ages and creates a signal that can be seen in absorption relative to the CMB. \citep{Furlanetto-Oh-Briggs, Pritchard-Loeb}.
%Absorption of the cosmic microwave background radiation by the hyperfine transition of hydrogen is not affected by complex astrophysics processes and would provide a clean probe of the universe’s evolution during the Dark Ages.
The 21\,cm signal traces the matter power spectrum throughout the Dark Ages, probing fluctuations all the way down to the Jeans scale; an all-sky, cosmic variance limited survey of the Dark Ages 21\,cm signal contains approximately one trillion times more modes than the CMB \citep{Loeb_2006, Furlanetto}.

However, there are enormous observational challenges to detecting this signal. Earth’s ionosphere refracts and absorbs low-frequency radio waves and, at the lowest frequencies, is completely opaque. 
%The redshifts of Dark Ages extend from $z \sim 1100$ to 50; the 21 cm signal would not actually be visible until $z \sim 150$ so the 21cm signal got reshifted from 1420 MHz to frequencies below 30 MHz. 
The observable period of the Dark Ages runs from approximately $z = 150$ (when the residual free-electron density in the hydrogen gas drops low enough to decouple it from the CMB) and $z = 50$ (when the first stars might conceivably start forming); this places the signal of interest in the range $10 - 30$\,MHz.

Such frequencies are almost impossible to observe from the ground due to the opaque ionosphere and human generated transmissions. There are nights (particularly during the solar minimum) when frequencies less than 30 MHz can be accessed from the ground, but reaching the required observing time ($\gtrsim 1000$ hours of observation; \citealt{Pober_2014}, Pober, \textit{in prep.}) with only a handful of nights per year during solar minimum makes it nearly impossible to observe the Dark Ages from Earth.

A space-based or a lunar-based radio telescope could enable 21\,cm signal from the Dark Ages to be detected \citep{Koopmans}. A radio array in space or on the far side of the Moon would allow us to avoid not the Earth’s ionosphere, but also terrestrial radio frequency interference (RFI) thanks to $>100\,\rm{dB}$ attenuation of RFI on the lunar far side. \citep{Farside}. The lunar far side also provides enough space to spread out a large number of antennas, which is required for the sensitivity and spatial resolution for the 21\,cm power spectrum analysis.

%Ground-based experiments are, however, currently working on addressing these issues for the detection of the 21 cm signal from the Epoch of Reionization. Due to anthropogenic radio frequency interference (RFI) from the Earth, it’s extremely challenging to study 21cm signal from the Earth, but there is $>100\,\rm{dB}$ attenuation of RFI on the lunar far side. 

However, even a lunar based interferometer will face significant challenges in pursuing a Dark Ages 21\,cm signal detection.  Foregrounds emission --- including supernovae remnants, pulsars, radio galaxies, and diffuse emission from the Milky Way ---  make it difficult to detect the redshifted 21\,cm signal at any frequency. As foregrounds are several orders of magnitude brighter than the 21 cm signal \citep{Santos}, extreme precision is needed to cleanly separate the two. %The main challenge is that techniques that remove foregrounds can also remove the signal. 

We can, in principle, differentiate the two because foregrounds are spectrally smooth and featureless, whereas the cosmological signal has features at different wavelengths.
A significant body of work has gone into studying foreground removal and/or mitigation for 21\,cm experiments. (see e.g. \citealt{Trott, Mertens, HERA} for examples of the latest techniques used by some of the leading experiments in the field.) 

However, there is a new potential foreground issue for very low frequency Dark Ages experiments: opacity of the foregrounds. Foregrounds are in front of the 21 cm signal, so if they are optically thick to low frequency radiation, they will obscure it entirely due to free-free absorption. The optical depth of this absorption is frequency-dependent, making it a potential problem for the low-frequency experiments studying the Dark Ages. The issue of foreground opacity presents both a theoretical and practical challenge for Dark Ages 21\,cm experiments.  First, if some fraction of the sky is optically thick, then the number of modes measurable are corresponding diminished.  Although an all-sky survey has the potential to provide vastly more information than the CMB, accurate forecasting will require us to know just how many modes we can actually hope to measure.  If there are frequencies where nearly the entire sky is optically thick, then the signal from the corresponding redshifts may be entirely unobservable.  Practically speaking, the first generation of 21\,cm Dark Ages experiments will aim for a detection of the signal, not the ``be-all and end-all'' survey of cosmology.  These experiments will likely target a small fraction of the sky free from bright and/or opaque foregrounds.  However, several leading ground-based experiments (e.g. HERA; \citealt{DeBoer_2017}) have eschewed steerable antennas for the sake of mechanical simplicity. Such a drift-scanning experiment is effectively at the whim of what foregrounds exist at the declination that passes through zenith.  For a lunar far side 21\,cm experiment, it therefore becomes important to know the \emph{spatial distribution} of opaque foregrounds regions, as that will determine the suitability of potential landing sites and may even make steerable antennas an essential mission requirement (driving up both cost and complexity).

At the lowest frequencies observed from the ground, a spectral turnover in the diffuse emission is already visible. The the goal of this paper is to use existing ground based data to infer what can be said about the opacity at even lower frequencies relevant to Dark Ages 21\,cm cosmology.  To find out which regions of the sky might be optically thick for the highest redshifts of the 21 cm signal, we fit the measurements from LWA1 \citep{Dowell} combined with the 408\,MHz Haslam map \citep{Haslam_1981,Haslam_1982,Remazeilles} with a two-component model from \citet{Cane}.  We then calculate the frequency-dependent foreground optical depth point-by-point across the sky. 

Studies of free-free absorption in the low-frequency radio sky have a long history. The Galactic plane, in particular, is known to be optically thick at frequencies of 20\,MHz and sometimes even higher \citep{Ellis&Hamilton, Cane&Whitham}.
At higher Galactic latitudes, the local densities of ionized hydrogen are much lower and free-free absorption does not appear to be significant for frequencies above $\sim2$\,MHz \citep{Reynolds}.
However, these older maps are generally low resolution and do not have well understood error properties.  (For the most part, they do not report any kind of error bar.)
%However, our model has some limitations in the galactic plane which we discuss further.} 
In our analysis, we focus on understanding the error properties of the measurements and report the statistical confidence with which we can say a particular region of sky is optically thick at a given frequency. We choose the LWA1 data because they provide uncertainty maps that enable us to accurately calculate the statistical significance of our results. This emphasis on statistical confidence sets our work apart from previous analyses (although our results are generally quite consistent with those studies).
%, gives us confidence in our foreground optical depth maps. }

The outline for the rest of this paper is as follows.  In \S\ref{sec:method}, we present both our model used to fit the existing observations and determine the optical depth (\S\ref{sec:canefit}) and the datasets we apply the model to (\S\ref{sec:data}). In \S\ref{sec:results}, we present the maps of the optically thick regions in the sky. In \S\ref{sec:discussion}, we discuss some of the sources of error, and finally in \S\ref{sec:conclusion} we present our conclusions.

\section{Method}
\label{sec:method}

Several studies have led to the creation of ``global sky models'' that seek to model and predict the radio foreground emission as a function of frequency across the entire sky \citep{GSM, GSM2016, Cong, eGSM}.
These projects combine information from a large number of radio surveys and must contend with differing sky-coverage, map resolutions, and error-properties.
%\footnote{Many low frequency sky maps, particularly older ones, do not report any errors at all.}
Because these global sky models effectively interpolate between existing data sets, their error properties can be quite complicated as well (particularly in terms of frequency-to-frequency correlation).
The recent work of \citet{Cong} is particularly relevant to the current study.  They develop a radio sky model from 10\,MHz to 1\,MHz which takes into account free-free absorption using models of the free-electron distribution in the Galaxy.

Our project is not so ambitious.  Rather than describe (and predict) the radio frequency emission across the entire sky at arbitrary frequencies, we seek to ask a single question: what is the optical depth of the foregrounds to the highly redshifted 21\,cm background? As such, we seek only simple functional fit to the low-frequency radio foreground spectrum (where the optical depth is one of the free parameters of the fit). This tailored approach allows us to focus on the statistical confidence, which will prove very important in interpreting the results. As we shall see, the data often prefers a fit that includes a spectral turnover, typically as as high as 10 MHz (e.g. the B pixel in Figure \ref{fig:pixelplot}). 
%but it would be a mistake to conclude so, as this result is not statistically significant. 
By focusing on error propagation, however, we can determine where that preference for a spectral turnover is actually statistically significant.

In this section, we first present the model from \cite{Cane} that we use to achieve this goal in \S\ref{sec:canefit}.  Then we discuss the datasets used in the analysis in \S\ref{sec:data}.

\subsection{A Model for the Spectrum of Low-Frequency Radio Foreground Emission}
\label{sec:canefit}

\cite{Cane} provides a simple model for the spectral behavior of the low-frequency radio foregrounds\footnote{The \cite{Cane} paper refers to Galactic emission as a ``background'' given the scientific interests of the time, but in this paper we will refer to it as a foreground.}, describing the average non-thermal foreground spectrum at high latitudes with two components. The non-thermal spectrum is a result of synchrotron radiation of cosmic ray electrons, which are rotating in the Galactic magnetic field. The turnover of the spectrum is caused by free-free absorption of the radiation by ionized hydrogen. The two-component model describes the spectrum as contributions from Galactic and extragalactic sources. The Galactic term leads to both emission and absorption. 

To confirm the validity of their model, \citet{Cane} measured the radio sky spectrum at many frequencies in the northern and southern hemispheres and combined their results with additional $100$ independent measurements by other observers. Although the model is over 40 years old, it remains an excellent description of the low-frequency radio sky. In more recent years, for example, \citet{Dulk} used it to calibrate the BIRS and WAVES experiments and it is cited as the reference model for the low-frequency sky temperature off the plane of the Galaxy in \citet{TMS} (TMS hereafter, see their equation 5.24).

This two component model is described by the following equation:
\begin{equation}
    I(\nu) = I_\mathrm{g}(\nu) \frac{1 - exp[-\tau(\nu)]}{\tau(\nu)} + I_\mathrm{eg}(\nu) exp[-\tau(\nu)],
\end{equation}
where $I_\mathrm{g}(\nu)$ and $I_\mathrm{eg}(\nu)$ are the Galactic brightness and extragalactic brightness in $\mathrm{W m^{-2} Hz^{-1} sr^{-1}} $ respectively, at frequency $\nu$ in MHz, and $\tau(\nu)$ is the frequency-dependent optical depth for absorption.  

An approximate relation for the optical depth of the free-free absorption by ionized hydrogen is:
\begin{equation}
    \tau(\nu) = 1.64 \times 10^5 T_{e}^{-1.35} \nu^{-2.1} \mathrm{EM},
\end{equation}
where $T_{e}$ is the electron temperature in K, \textbf{$\mathrm{EM}$} %\footnote{The \cite{Cane} paper refers to the emission measure as "E", but it is more common to use "EM" so that's how we defined it.}
is the emission measure in $\mathrm{ cm^{-6} pc}$ and $\nu$ is frequency in MHz. This approximation was originally found by \citet{Mezger}. To simplify this relation, we can combine all the frequency independent factors into a single ``optical depth coefficient'', $F$, as:
\begin{equation}
   \tau(\nu) = F \nu^{-2.1} 
\end{equation} 
We note that this approximation is under the assumption that all ionized regions along the line of sight have the same temperature, which may not be correct at the lowest Galactic latitudes.

We can also express frequency dependence of the specific intensity in terms of spectral indices from the Galactic ($\alpha_1$) and extragalactic ($\alpha_2$) sources as:
\textbf{$I_\mathrm{g}(\nu) = I_\mathrm{o,g} \nu^{-\alpha_1}$} and $I_\mathrm{eg}(\nu) = I_\mathrm{o,eg} \nu^{-\alpha_2}$.
 We assume that extragalactic parameters $\alpha_2$ and $I_\mathrm{eg}$ are constant across the entire sky and can be treated as fixed numbers. We take $\alpha_2$ to be $0.8$, based on observations by \citet{Simon} and also the value used in \citet{TMS}. We take $I_\mathrm{eg}$ to be $5.5\times10^4\,\rm{K}$ as obtained by \cite{Cane}. Our approach is to fit for the remaining three parameters $I_\mathrm{o,g}$, $\alpha_1$, and $F$ using the data described in \S\ref{sec:data}. 
 
 We proceed by fitting multifrequency radio sky maps pixel-by-pixel to determine the spatial distribution of the three Galactic parameters in Cane's two-component model using a nonlinear least squares fit. For our nonlinear least squares fit we used the `trf' (Trust Region Reflective) \citep{TRF} algorithm implemented in \texttt{scipy.optimize.curve\_fit} \footnote{\url{https://docs.scipy.org/doc/scipy/reference/generated/scipy.optimize.least_squares.html}}, which is motivated by solving a system of equations with the first-order optimality condition for a bound-constrained minimization problem. It iteratively solves subproblems, which are adjusted by a special diagonal quadratic term, in the trust region shape, which is determined by the direction of the gradient and the distance from the bounds. The bounds used to find $I_\mathrm{o,g}$ were $[0.5, 8000]$ $\times 10^{-20} \mathrm{W m^{-2} Hz^{-1} sr^{-1}} $, $\alpha_1$ were $[0.1, 2]$ and $F$ were \textbf{$[0, 12000]$}, respectively.
 This prevents the algorithm from making steps directly into the bounds and allows it to explore the whole space of variables. 
These bounds effectively serve as hard priors on the parameters of our fits, but they span a sufficiently broad range so as to not affect our results.
 The algorithm also calculates the covariance of the fitted parameters by minimizing the sum of squared residuals over the input errors of the data points. %which were provided by the LWA1 accompanying error maps that will be described in more detail in the next section. 

 \subsection{Datasets}
 \label{sec:data}
 We use two principal data sets in this work: the LWA1 Low Frequency Sky Survey \citep{Dowell} maps and the all-sky Haslam map at 408\,MHz \citep{Haslam_1981,  Haslam_1982,Remazeilles}.  We describe each of these datasets in turn. In short, though, our motivation for focusing on these two data sets are (i) their significant sky coverage, (ii) well-understood systematics (including zero-point correction), and (iii) available uncertainty estimates.\footnote{Even though the Haslam map does not provide uncertainties, its role as a key foreground map for CMB studies has led to continued refinement of the data products \citep{Remazeilles} and external studies of its error properties (\citealt{Dowell}, Kim et al., \textit{in prep.})}
 
The LWA1 \citep{Taylor, Ellingson} is located in New Mexico, USA, and it consists of 256 dual-polarization dipole antennas. LWA1 was used to conduct the Low Frequency Sky Survey, which spans over the 10–88 MHz frequency range. The dataset contains the NSIDE 256 HEALPix \citep{Healpix} maps of brightness temperature at 35, 38, 40, 45, 50, 60, 70, 74, 80 MHz all with a bandwidth of 957 kHz and spatial resolutions from $4.7^{\circ}$ to $2^{\circ}$, respectively. The angular size of an NSIDE 256 pixel is $0.229$ degrees. In addition, each map has a corresponding uncertainty map that was generated using the mosaicking method by \citet{Dowell}. These uncertainty maps contain the effects of  uncertainties in the calibrations and other corrections that were applied to the data. The LWA1 maps were converted from intensity to temperature using the two-dimensional beam area \citep{Dowell}. Critically for our study, they correct for the missing Galactic emission on the largest spatial scales in the interferometric maps using the data from LEDA-64 New Mexico Deployment \citep{Taylor}. They also used forward modelling of LWA1 to determine which spatial scales are missing in the images. LWA1 is an ideal telescope for our study because its primary motivation is to provide a detailed picture of foregrounds for 21 cm cosmology at low frequencies. LWA's combination of observatory latitude, sensitivity and degree of aperture filling make sure that the images can be used for studying the spectral structure of foreground emission.

 In addition to the LWA1 maps, we also include the 408\,MHz Haslam map \citep{Haslam_1981,  Haslam_1982} to better constrain our model at higher frequencies. This total-power radio survey covers the entire sky at $1^{\circ}$ resolution. We chose the desourced and destriped Haslam map at 408 MHz \citep{Remazeilles}, which has removed the brightest sources $\geq 2$\, Jy and provides significant improvement to the original map. As the Haslam map does not provide error bars, we assumed $10\%$ error bars on each pixel, which was used by  \citet{Dowell} for the LWA1 Sky Model  and was found to be an appropriate level in the forthcoming extended Global Sky Model (eGSM; Adrian Liu, private communication). In addition, we added a zero-level correction, which is a second order correction to the calibrated data describe in \citet{Guzman}. It is described in the equation below:
 \begin{equation}
   T_{\nu,0} = T_\mathrm{CMB} + T_{\nu,\mathrm{Ex}} + T_{\nu,\mathrm{ZLC}}
\end{equation}
where $T_\mathrm{CMB}$ is Cosmic Microwave Background temperature, $T_{\nu,\mathrm{Ex}}$ is the extragalactic non-thermal background temperature and $T_{\nu,\mathrm{ZLC}}$ is Zero-level correction temperature.

The Haslam map is NSIDE 512 HEALPix map, which we downgraded to NSIDE 256 to match the LWA1 maps. Our uncertainty map is also calculated using 10\% of the brightness in each pixel of the lower resolution map of Haslam.

\section{Results}
\label{sec:results}

Using the Eq. 1 we performed a pixel-by-pixel fit to get HEALPix maps of $F$, $I_\mathrm{o,g}$, $\alpha_1$ at the NSIDE 256 resolution of the LWA1 maps, which are shown on the left in Fig. 1. We plot the corresponding standard deviations at each pixel on the right in Fig. 1.

\begin{figure*}[!tbp]
  \centering
  \includegraphics[width=0.5\columnwidth]{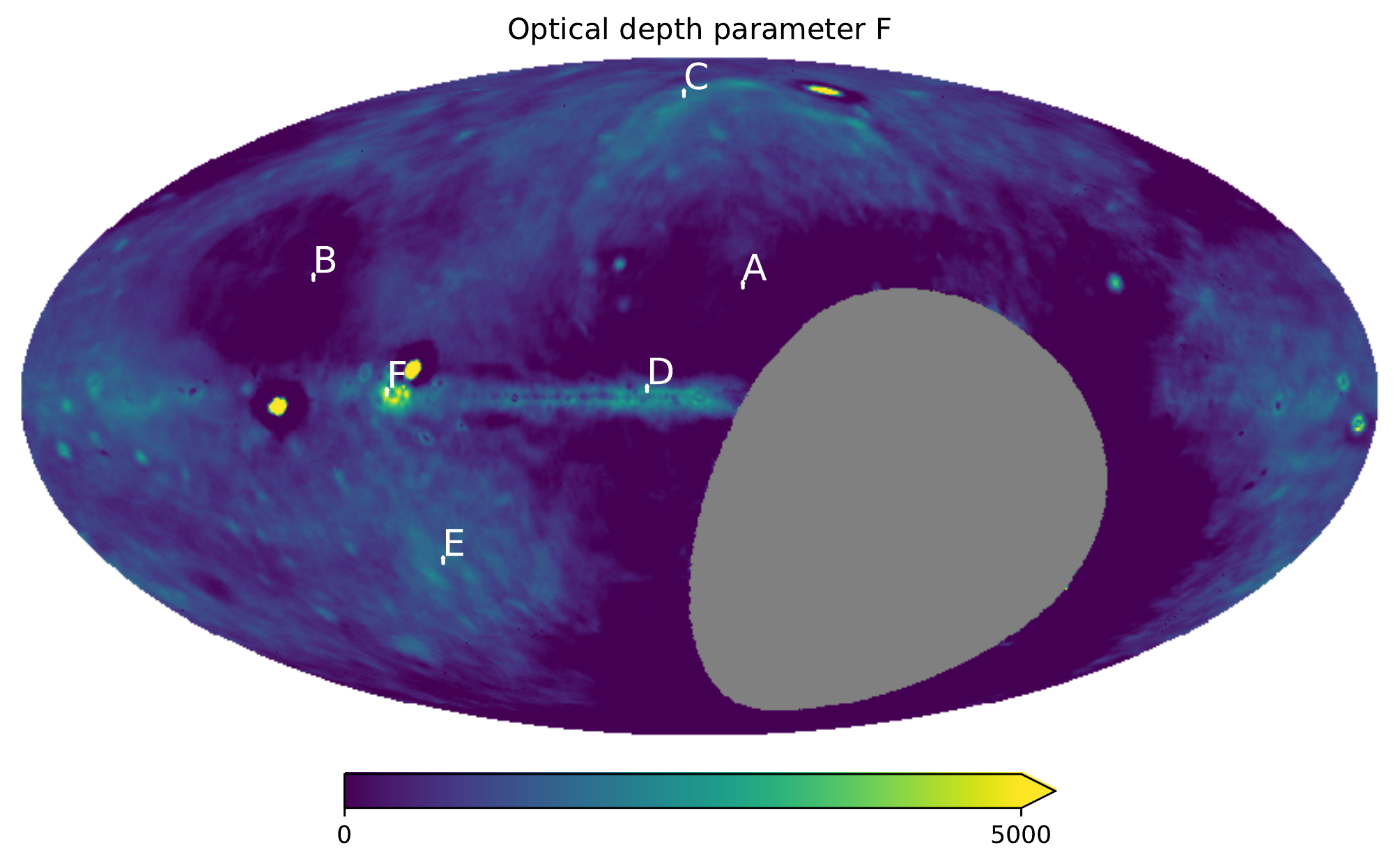}\includegraphics[width=0.5\columnwidth]{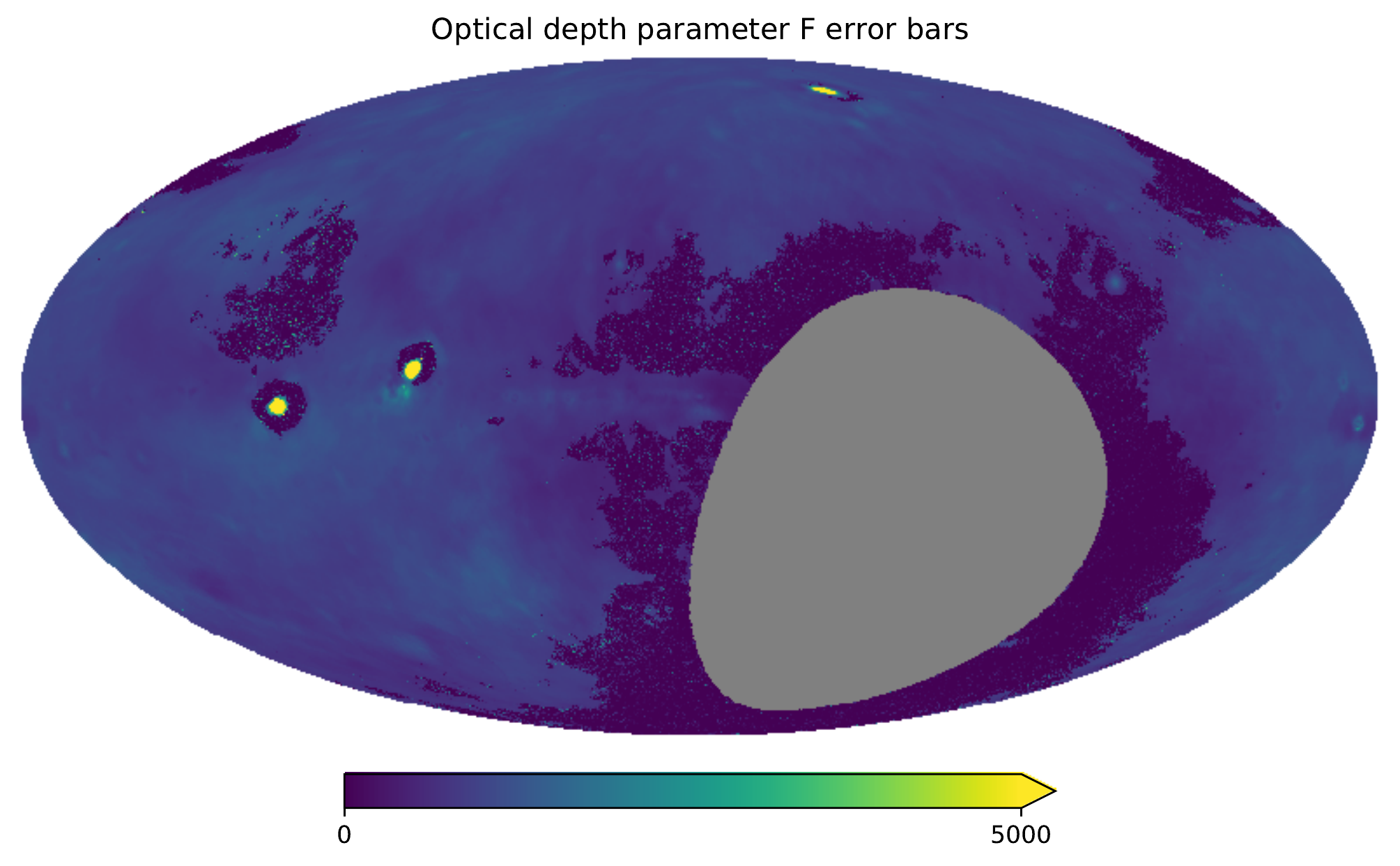}
  \label{fig:f1}
  \includegraphics[width=0.5\columnwidth]{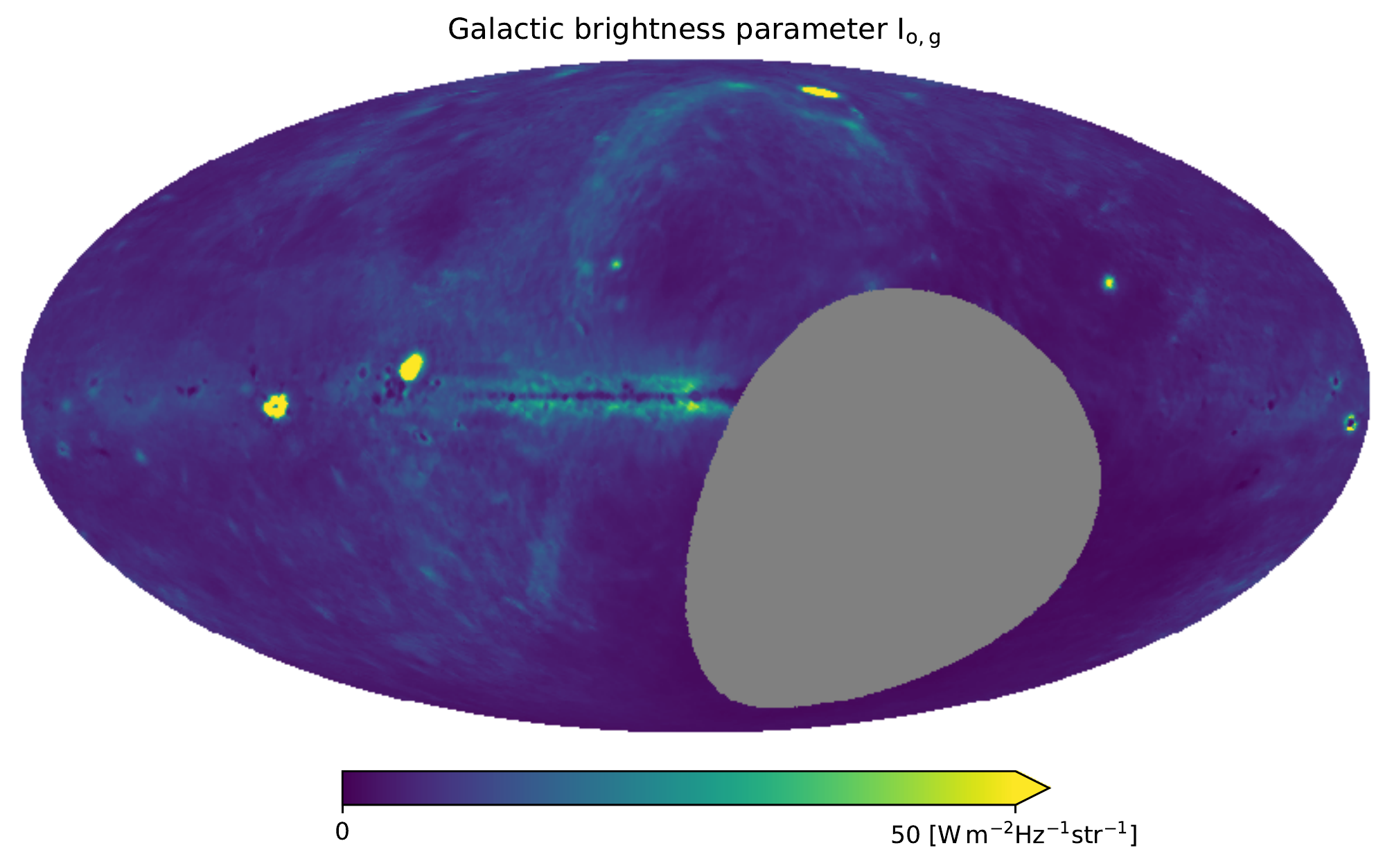}\includegraphics[width=0.5\columnwidth]{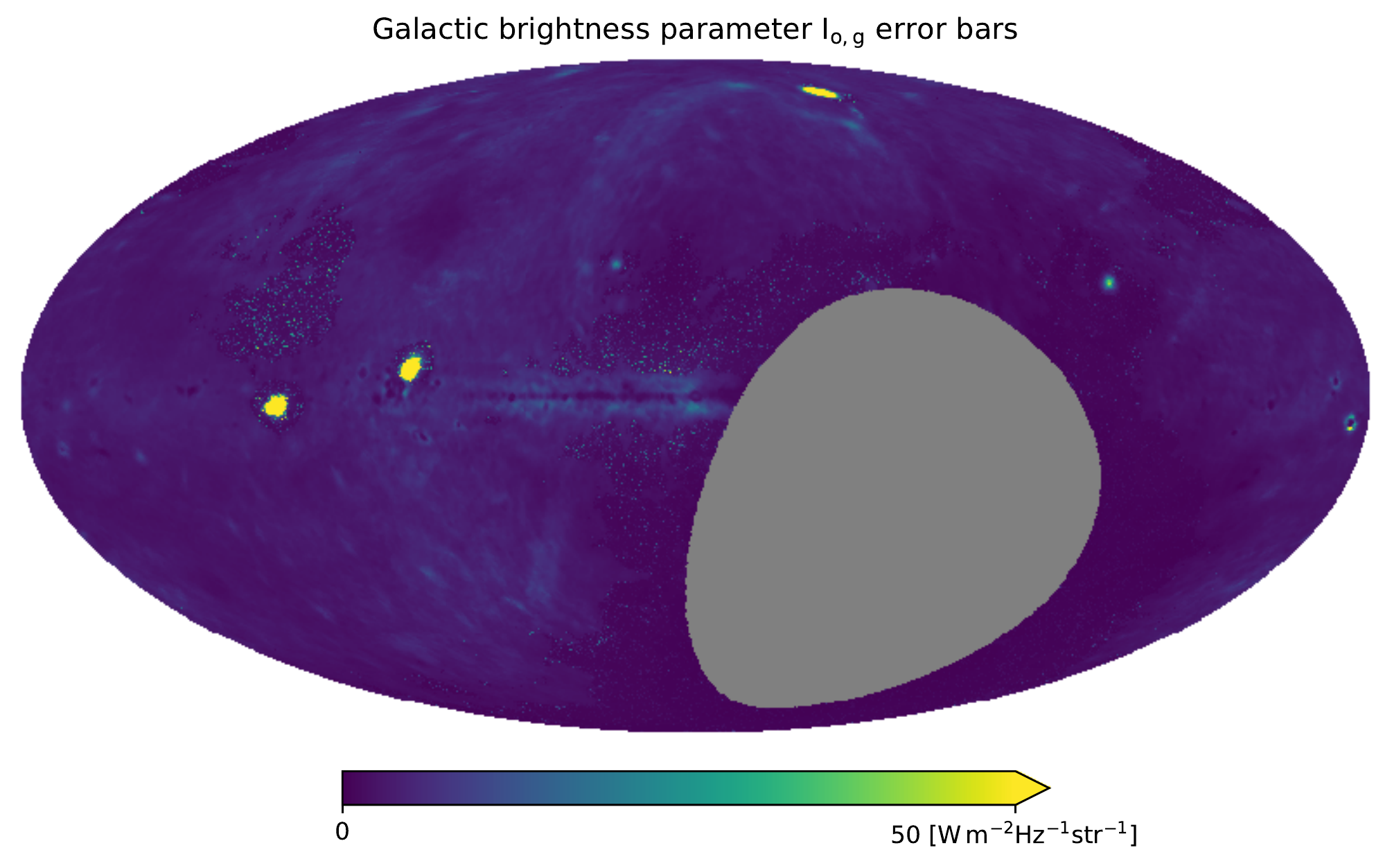}
  \label{fig:f2}
  \includegraphics[width=0.5\columnwidth]{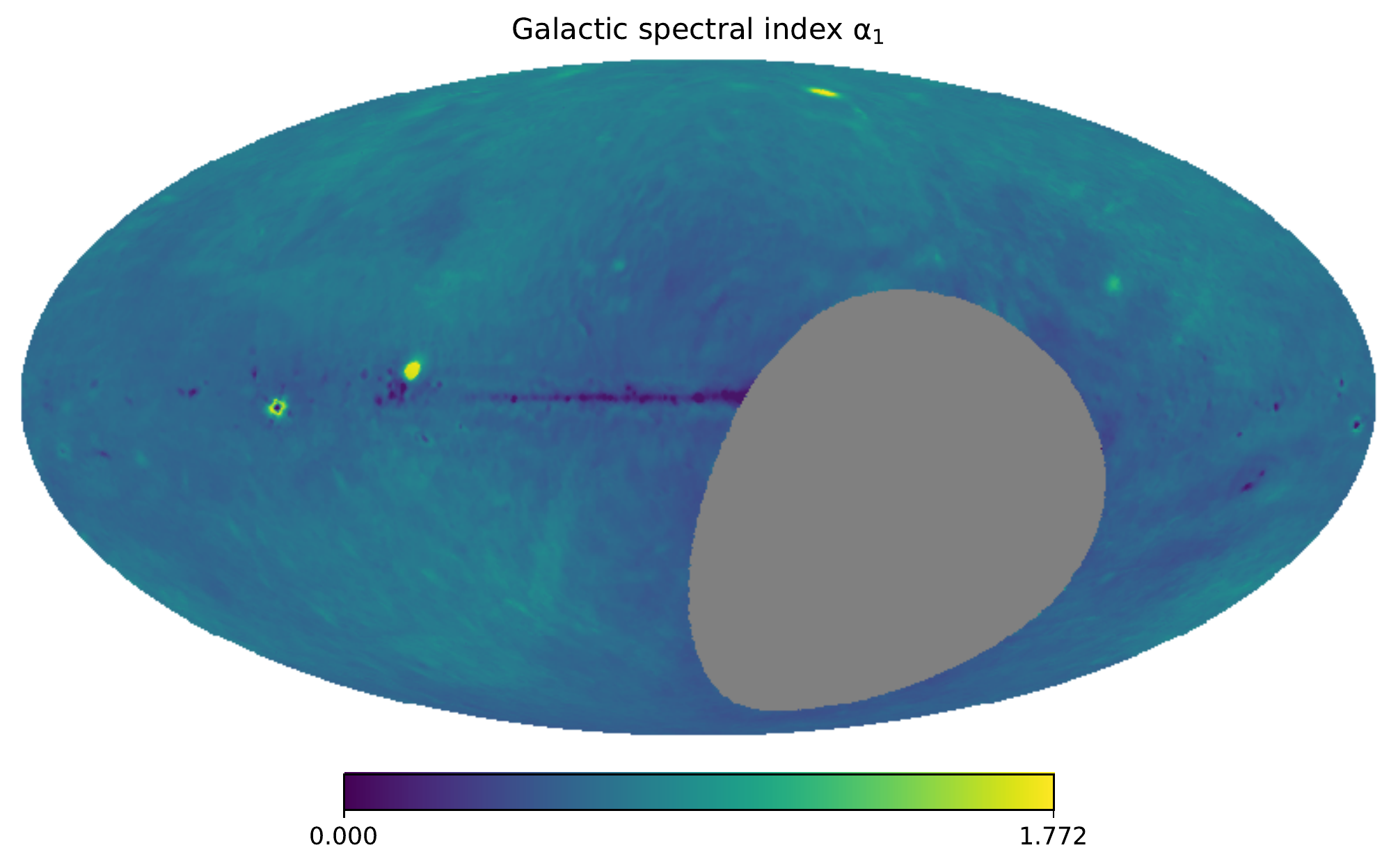}\includegraphics[width=0.5\columnwidth]{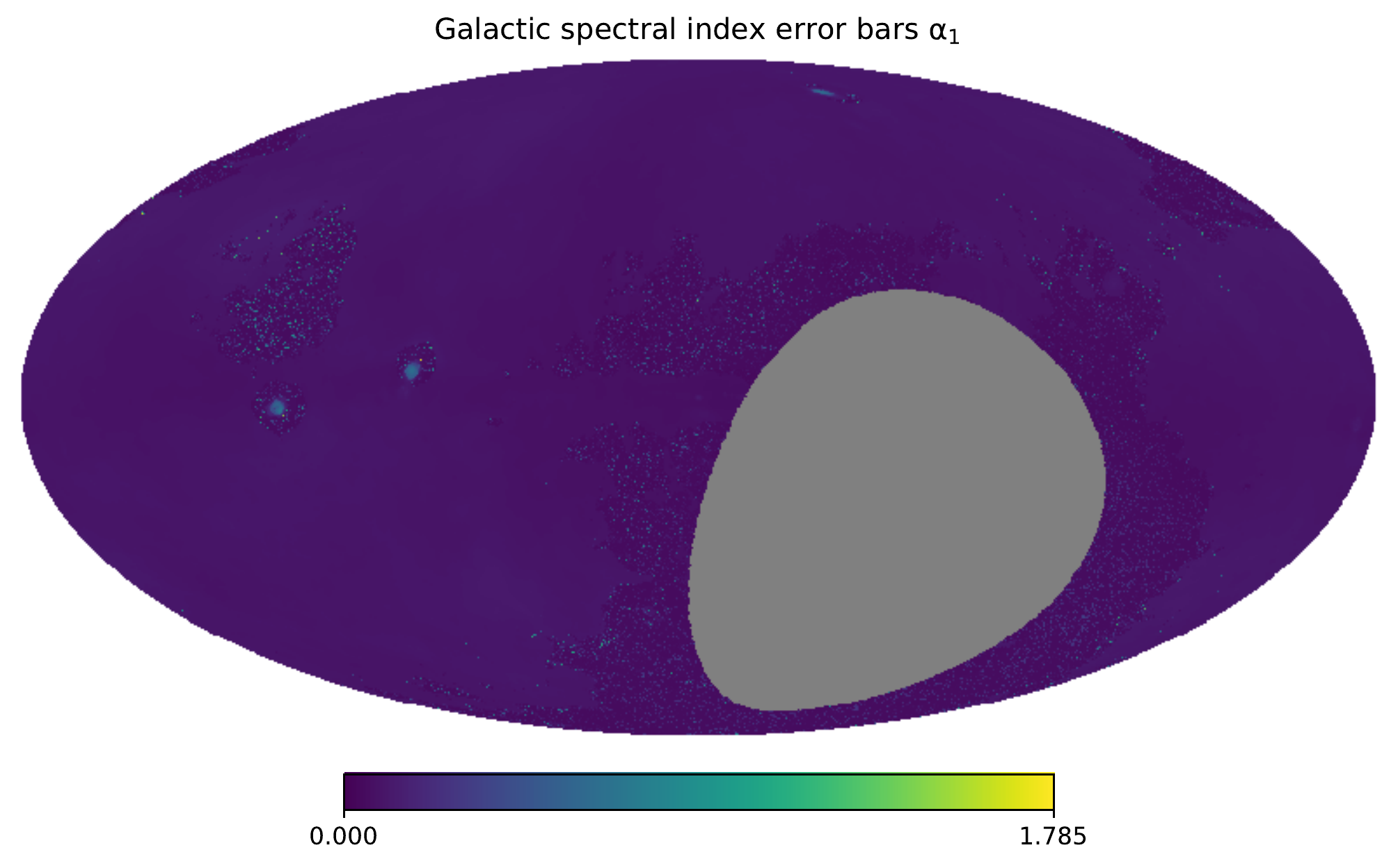}
  \label{fig:f3}
  \caption{ \textit{Left}: HEALPix maps of the values of $F$, $I_\mathrm{o,g}$ and $\alpha_1$, from top to bottom, from our fits. The letters in the upper left map correspond to pixels whose spectra are plotted in Figure 2. \textit{Right}: The corresponding standard deviations at each pixel for each of our three fit parameters. All six maps are in Galactic coordinates, with Galactic center at the middle of the maps.}
  \label{fig:fit_maps}
\end{figure*}

To illustrate how these fits are behaving we selected several pixels which span a range of environments and are broadly representative of the different kinds of spectral behavior seen in the data. The pixels we chose are indicated in the upper left panel of Figure 1 with letters. Pixels A, B, and E are chosen to be at high Galactic latitudes, far away from known regions of emission.  Pixels D and F, on the other hand, are well within the Galactic plane.  Pixel C is at high latitudes, but in the North Polar Spur of the Galaxy.  The measured spectrum for each of these pixels (blue points) and our fit (red line) is shown in Figure 2. The vertical dashed line shows the frequency at which the optical depth reaches unity (see discussion below) and the shaded gray regions show the 1, 2, and 3$\sigma$ equivalent uncertainties on this value. We see that our model provides a good fit to the data across all the different conditions probed.  We discuss the calculation of the confidence intervals below, but we can see several general classes of fits: Pixel A shows no evidence for a spectral turnover; the fit to Pixel B prefers a spectral turnover, but is of very weak ($<1\sigma$) statistical significance; Pixels C, D, and E all show marginal evidence for a spectral turnover ($>1\sigma$ but $<2\sigma$); and Pixel F shows relatively convincing evidence of a turnover ($>2\sigma$ but $<3\sigma$).  %The shaded grey regions are uncertainty bands and the calculation of the error is described below. 

\begin{figure*}[!tbp]
  \centering
  \includegraphics[scale=0.6]{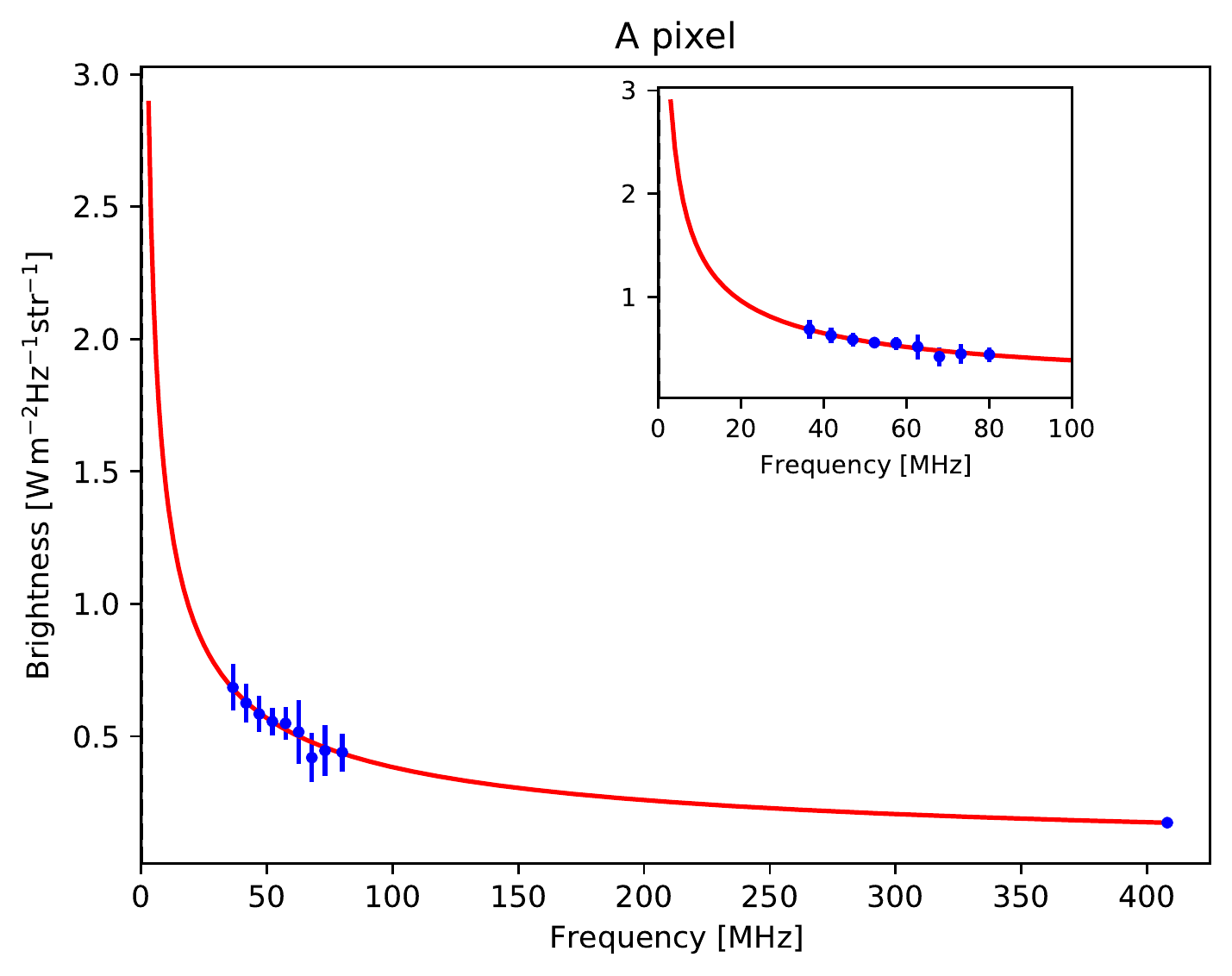}\label{fig:f1}
  \includegraphics[scale=0.6]{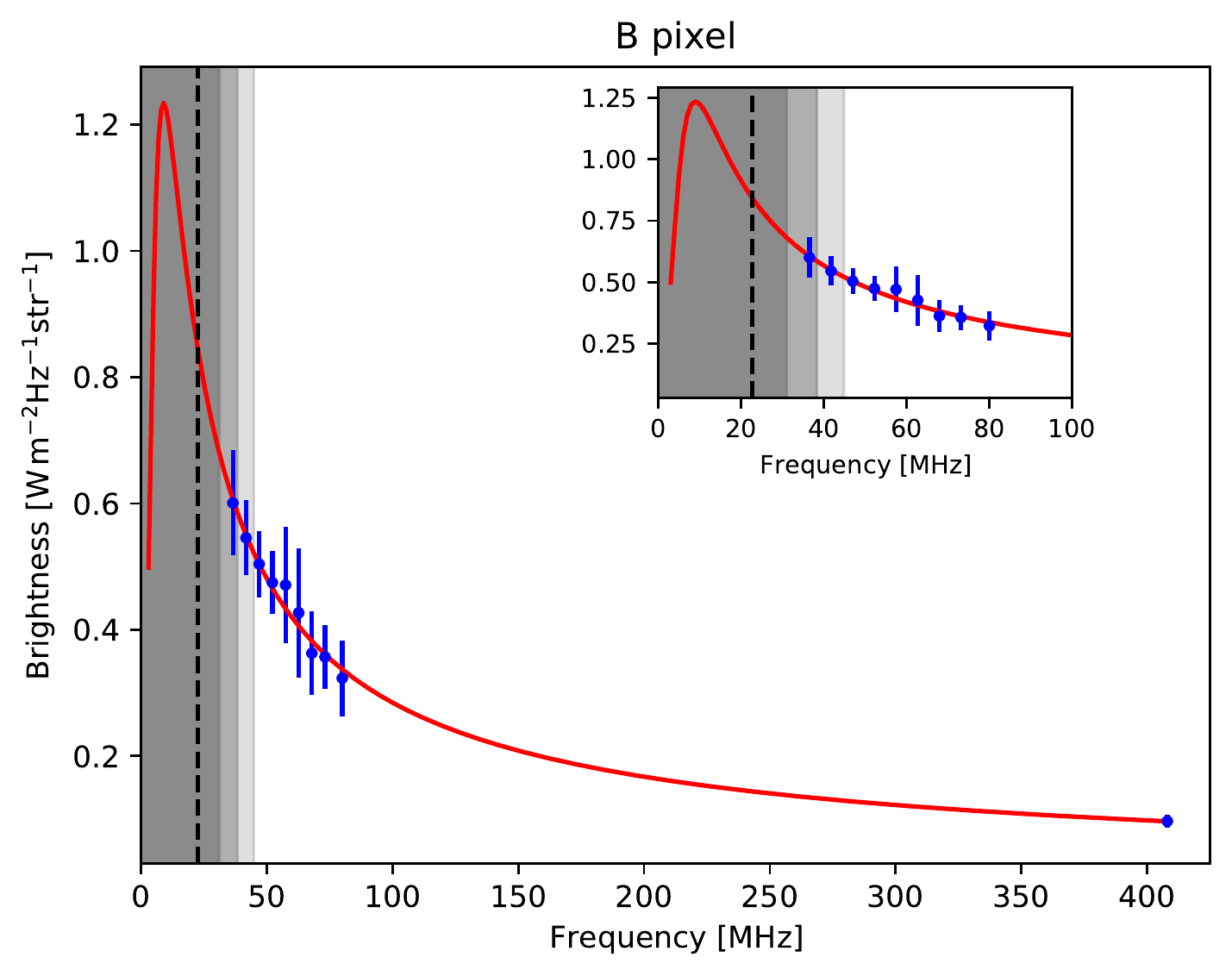}\label{fig:f2}
  \includegraphics[scale=0.6]{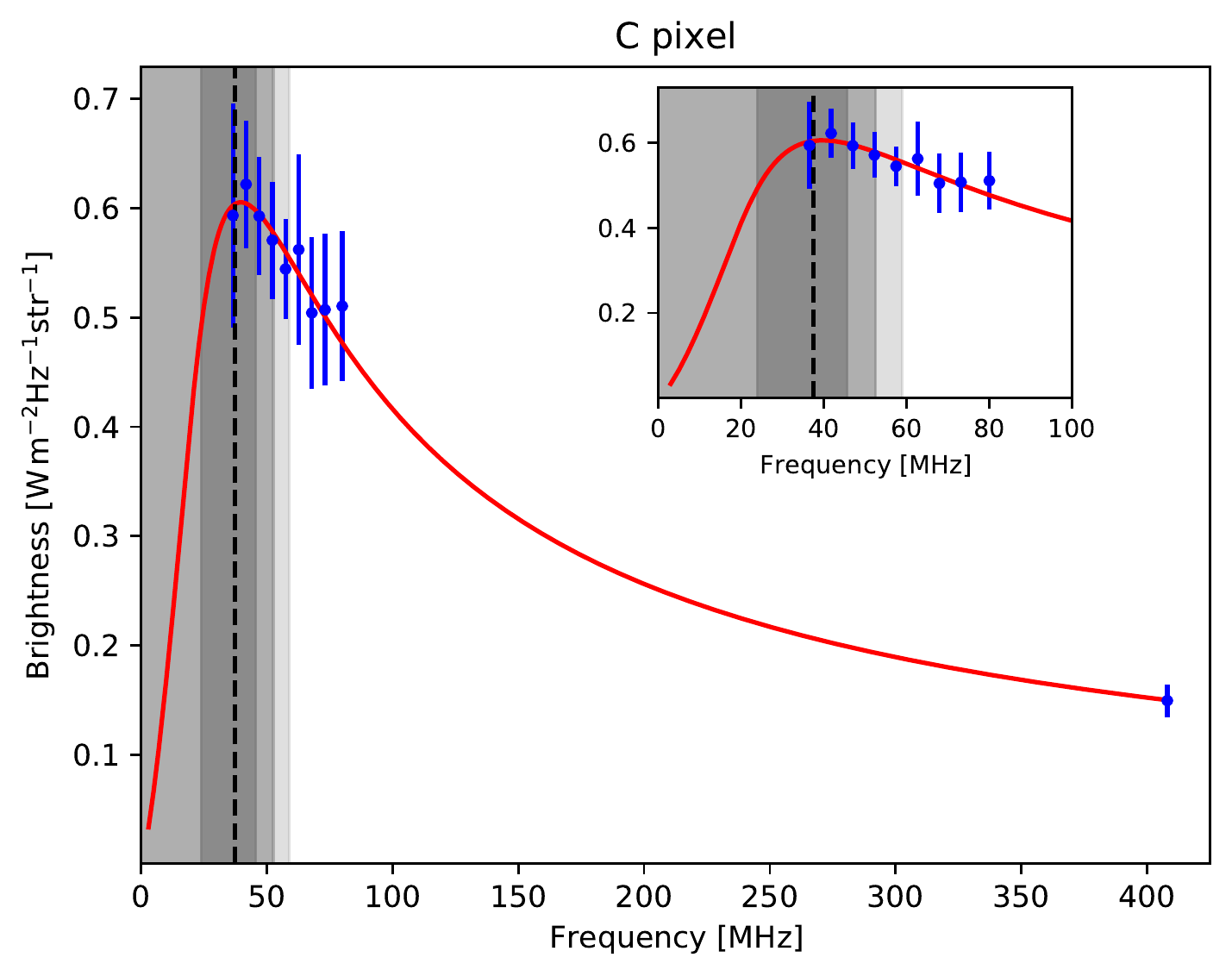}\label{fig:f3}
  \includegraphics[scale=0.6]{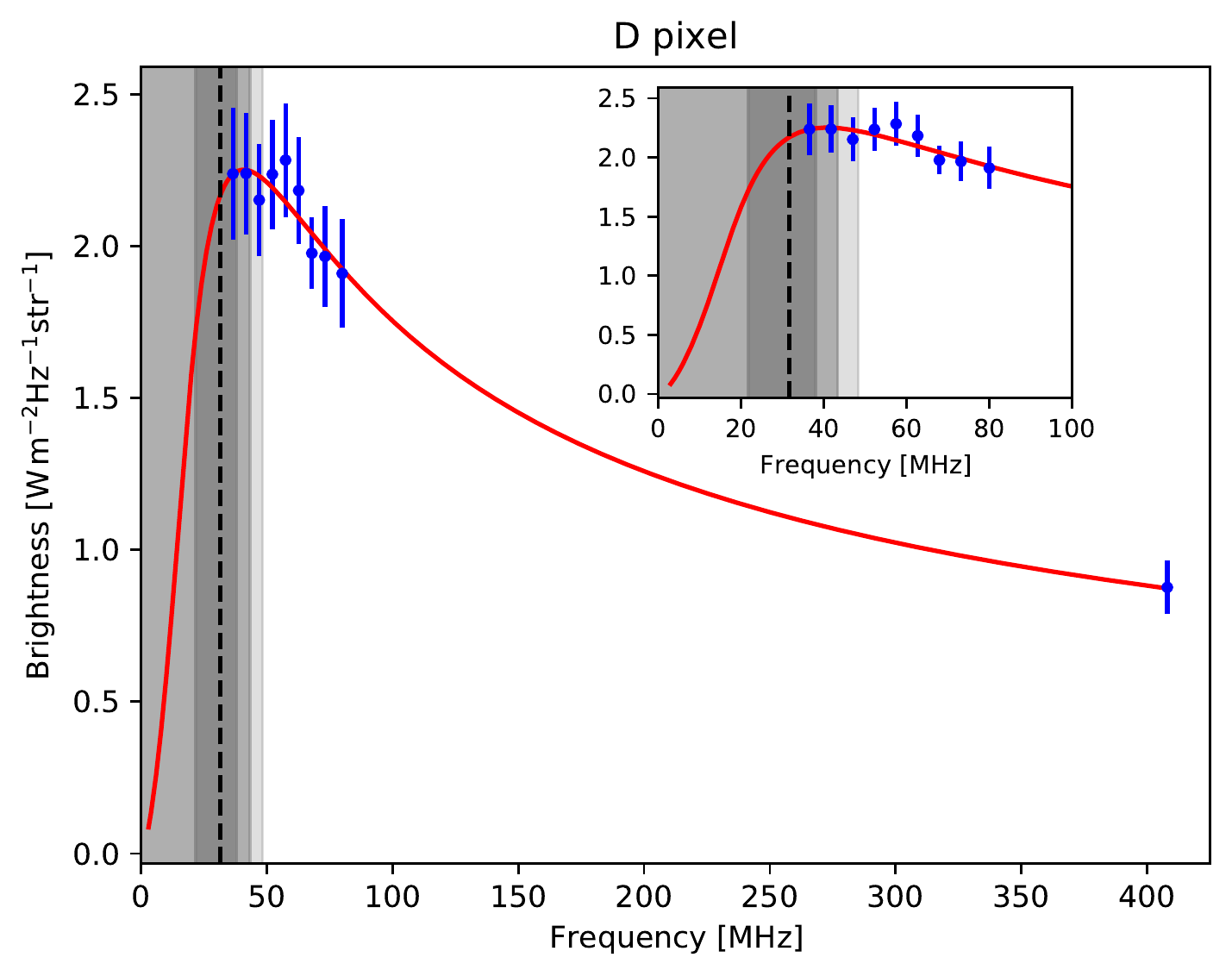}\label{fig:f4}
  \includegraphics[scale=0.6]{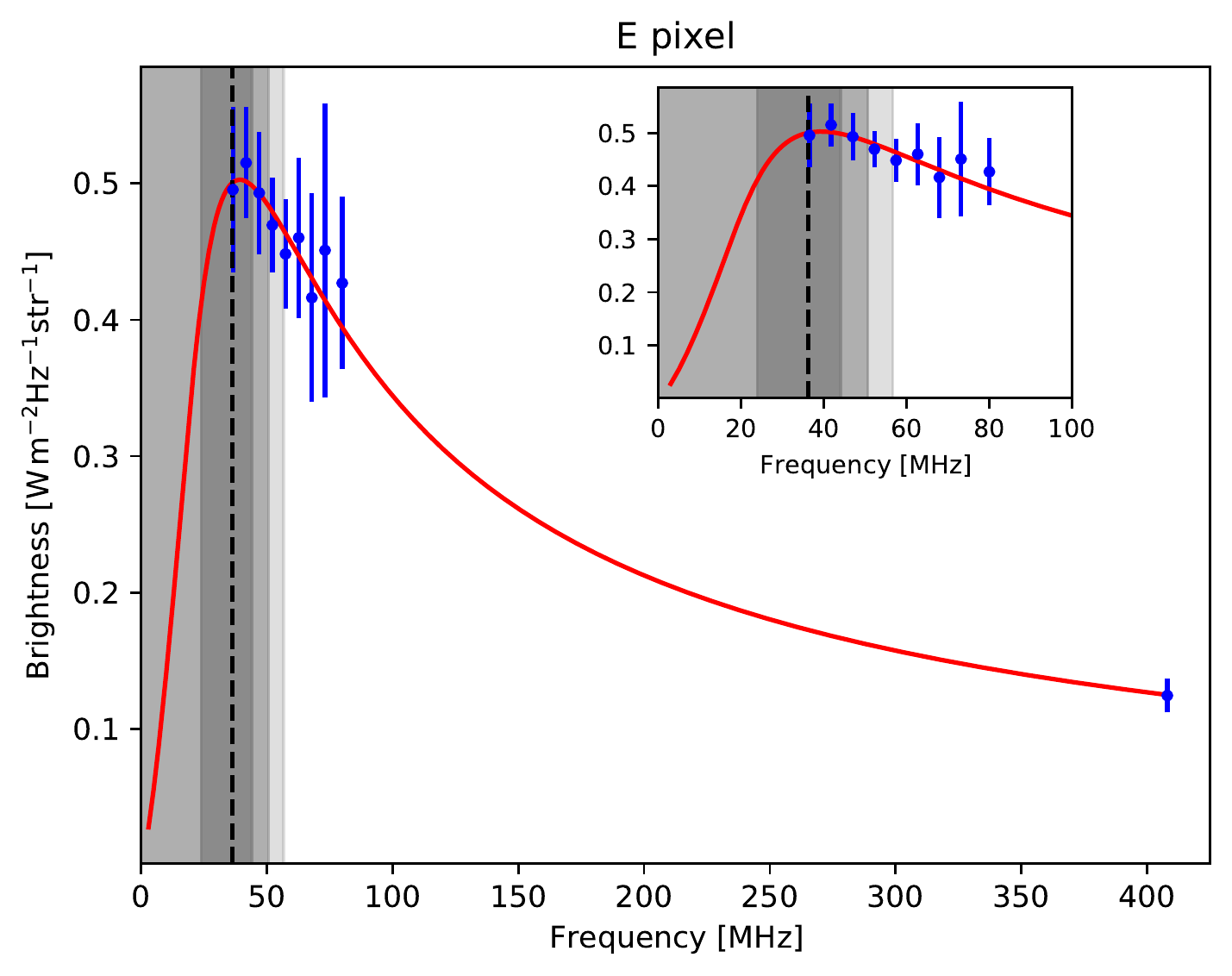}\label{fig:f5}
  \includegraphics[scale=0.6]{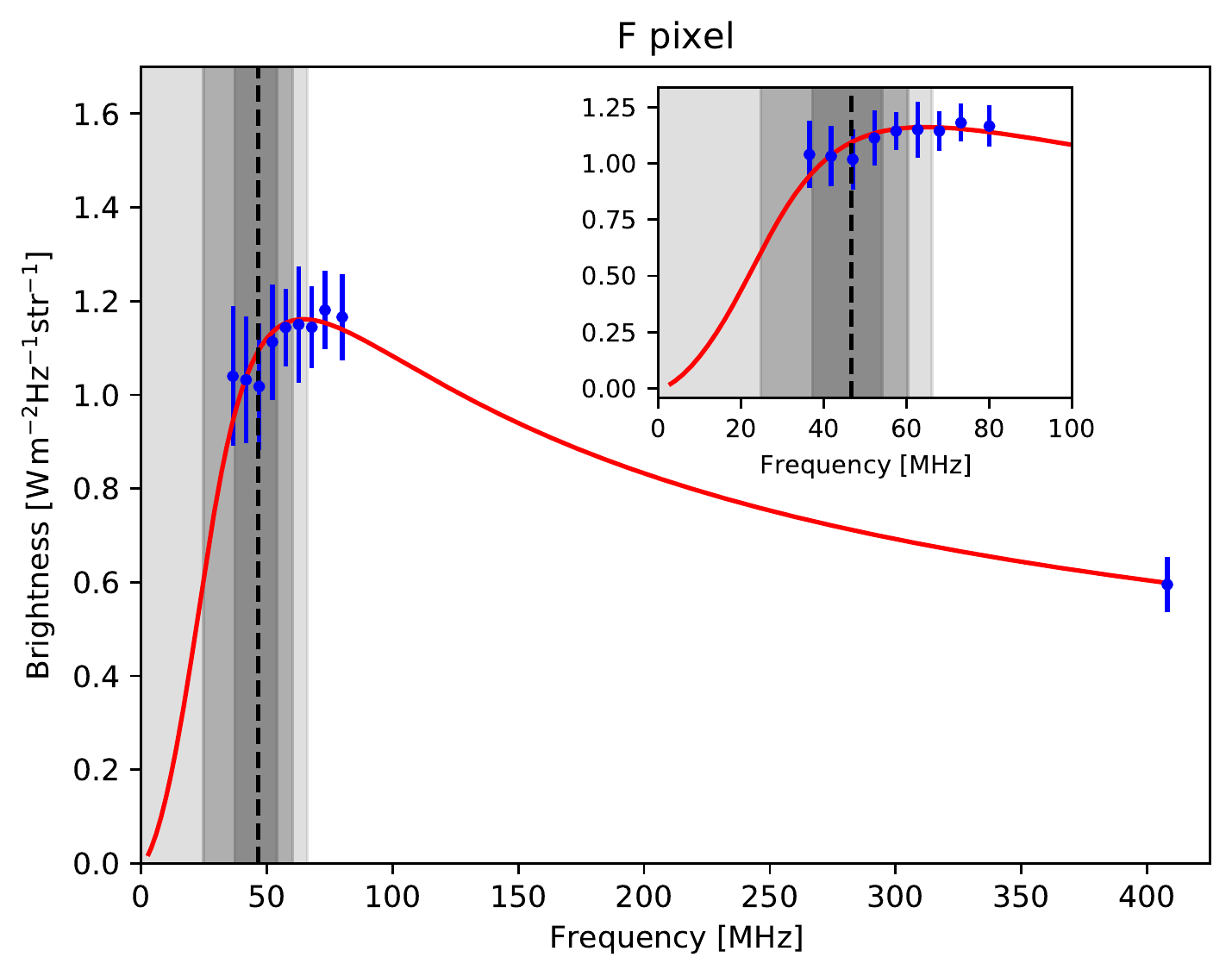}\label{fig:f6}
  \caption{ The measured data values (blue points) and fits (red lines) for the six pixels indicated in the upper left panel of Figure 1.  The vertical dashed line shows the frequency at which the optical depth reaches unity and the shaded gray regions from darkest to lightest show the 1, 2, and 3$\sigma$ equivalent uncertainties on this value, respectively (see text for details).Pixel A shows no evidence for a spectral turnover, and so the shaded grey regions are all effectively at zero and are not visible on the plot. The inset subplots in each panel zoom in on the frequency range $0-100$\,MHz.} %As we can see all the plots show a turnover but some pixels have turnover close to 0 so to high statistical significance they don't have a turnover. The shaded grey regions are uncertainty bands and their calculation is described in more detail in Section 3.} 
  \label{fig:pixelplot}
\end{figure*}

From the map of the optical depth coefficient, we need to determine which areas of the sky are optically thick to various redshifts of the 21\,cm signal. In order to do that, we find the frequency where the optical depth $\tau = 1$ and expressed Eq. 3 in the following way:
\begin{equation}
    \tau = F \nu^{-2.1} = 1
\end{equation}
which we rewrote to find the expression for the frequency at which it becomes optically thick:
\begin{equation}
\label{eq:F_to_nu}
    \nu = F^{1/2.1} 
\end{equation}

To determine the uncertainty on the frequency at which a pixel becomes optically thick, we used Monte Carlo simulation. Using the uncertainties we calculated from our fit, we create one million possible optical depth coefficients normally distributed around the best fit value. Each value in the distribution is then converted to frequency using Eq. \ref{eq:F_to_nu}. We then calculated the mean and ``1, 2 and 3 sigma equivalent'' widths of that distribution, i.e. the distances from the left and the right off the peak of the distribution which contain $68.2\%$, $95\%$  and $99.7\%$ of the probability, respectively. We express our results in terms of the statistical confidence with which we can claim a pixel to be optically thick (to 21\,cm emission from a particular redshift).  For example, consider a pixel where, using Eq. \ref{eq:F_to_nu}, we determine that the the sky becomes optically thick at 15\,MHz.  Using our Monte Carlo simulations, we then determine the uncertainties on that frequency to be $\pm 3$\,MHz ($1\sigma$-equivalent) and $\pm 8$\,MHz ($2\sigma$-equivalent).\footnote{Note that because we use Monte Carlo simulations to calculate the error distribution, the $2\sigma$-equivalent error need not equal twice the $1\sigma$-equivalent error (as in this example, but is generally true of our results), nor do the errors need to be symmetric around the mean (which is not the case in this example).}
We then compare with 9.4\,MHz, the frequency of redshifted 21\,cm emission from $z=150$. In this case, the pixel is said to be optically thick to 21\,cm emission from $z=150$ at greater than $1\sigma$ confidence (but not at greater than $2\sigma$ confidence).

Once we found the turnover frequency and the corresponding error bars, we created HEALPix maps of the sky to show which regions are optically thick for 21\,cm signal. As the optical depth is frequency dependent, we needed to pick a specific frequency to determine whether it is optically thick.  We chose frequencies corresponding to several redshifts of interest for 21\,cm cosmology: $z=50$, 100, and 150. Figure 3 shows which pixels are optical thick at each of these three redshifts and the associated confidence.

\begin{figure*}[!tbp]
  \centering
  \includegraphics[scale=0.6]{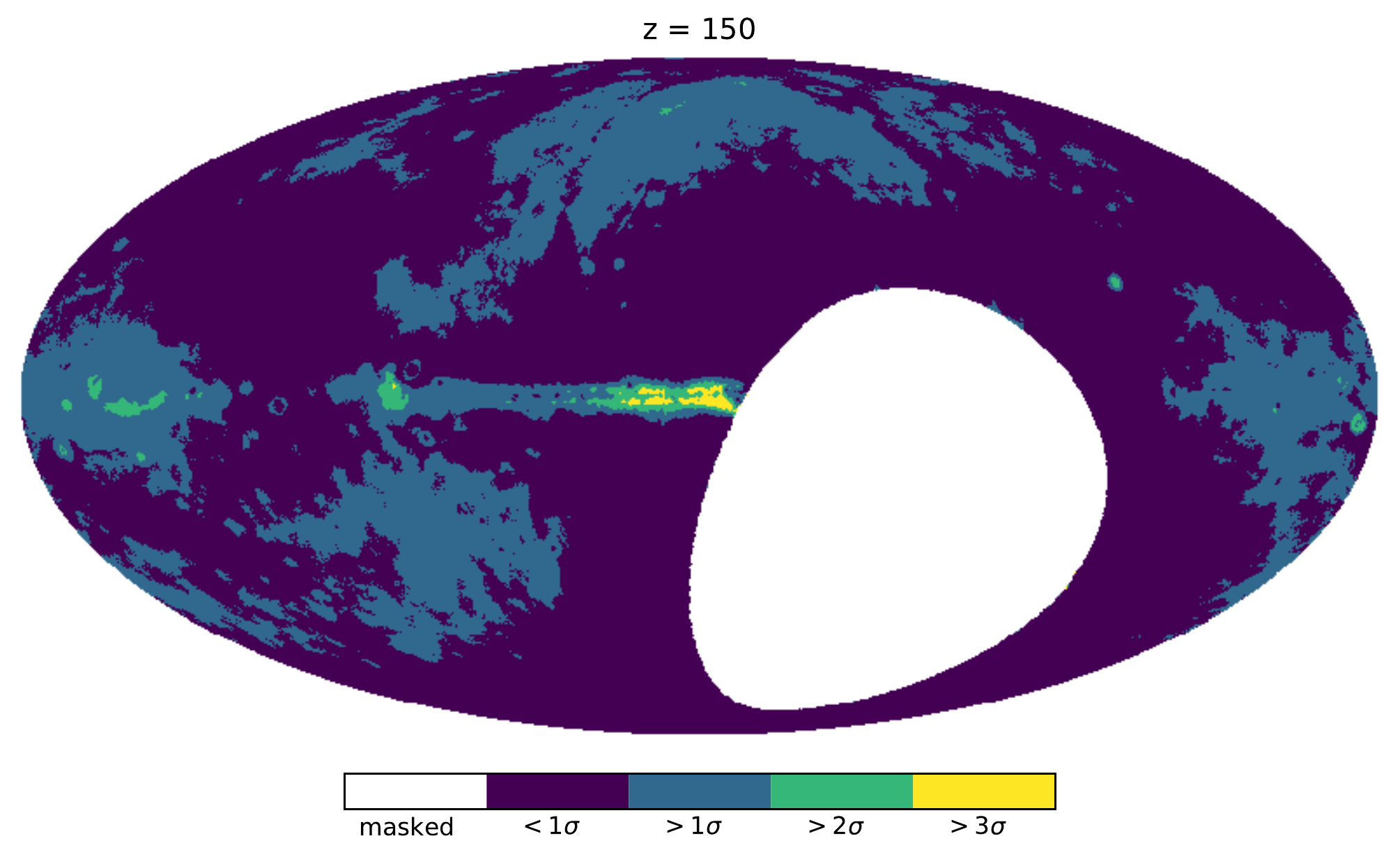}\label{fig:f1}
  \hfill
  \includegraphics[scale=0.6]{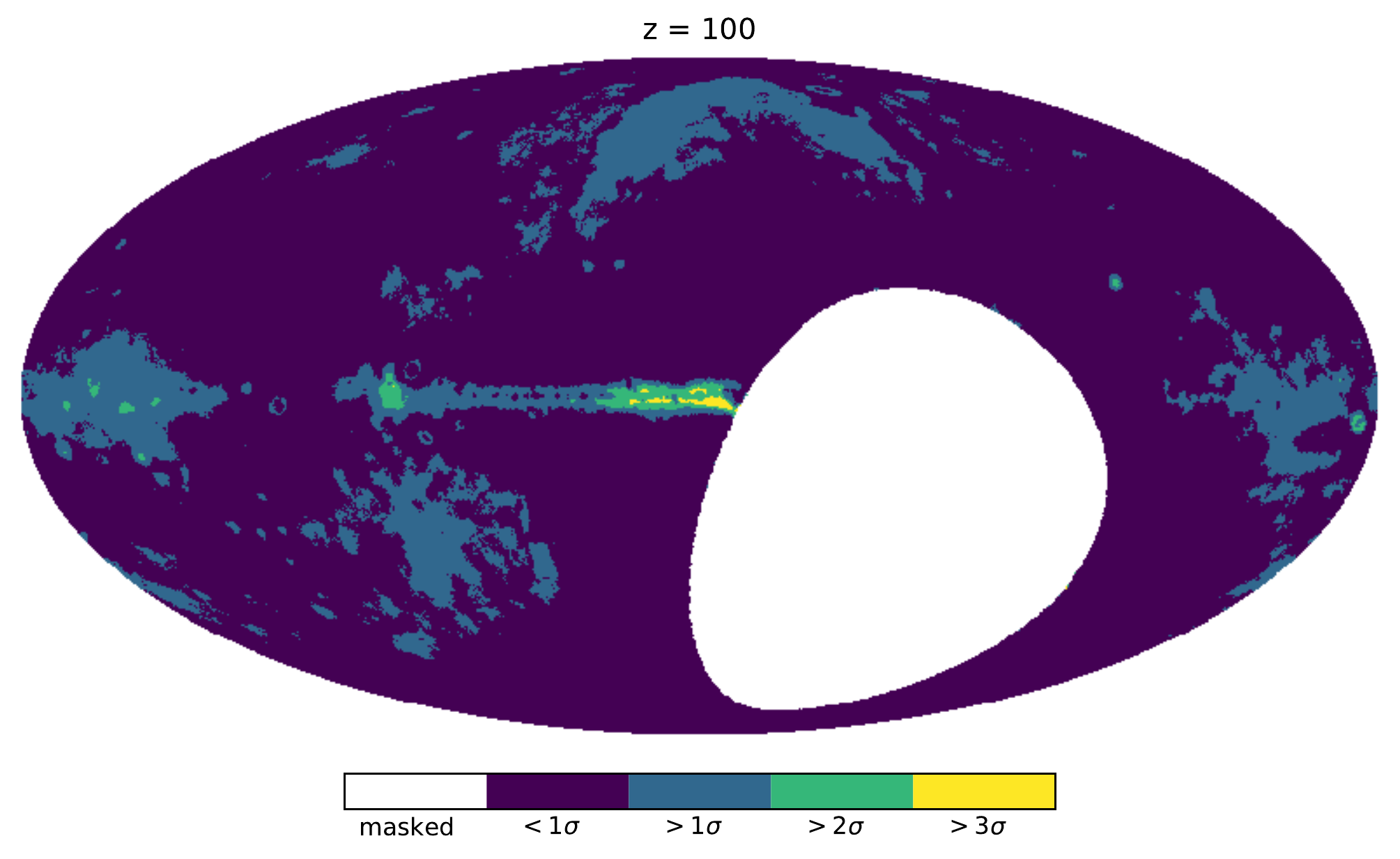}\label{fig:f2}
  \includegraphics[scale=0.6]{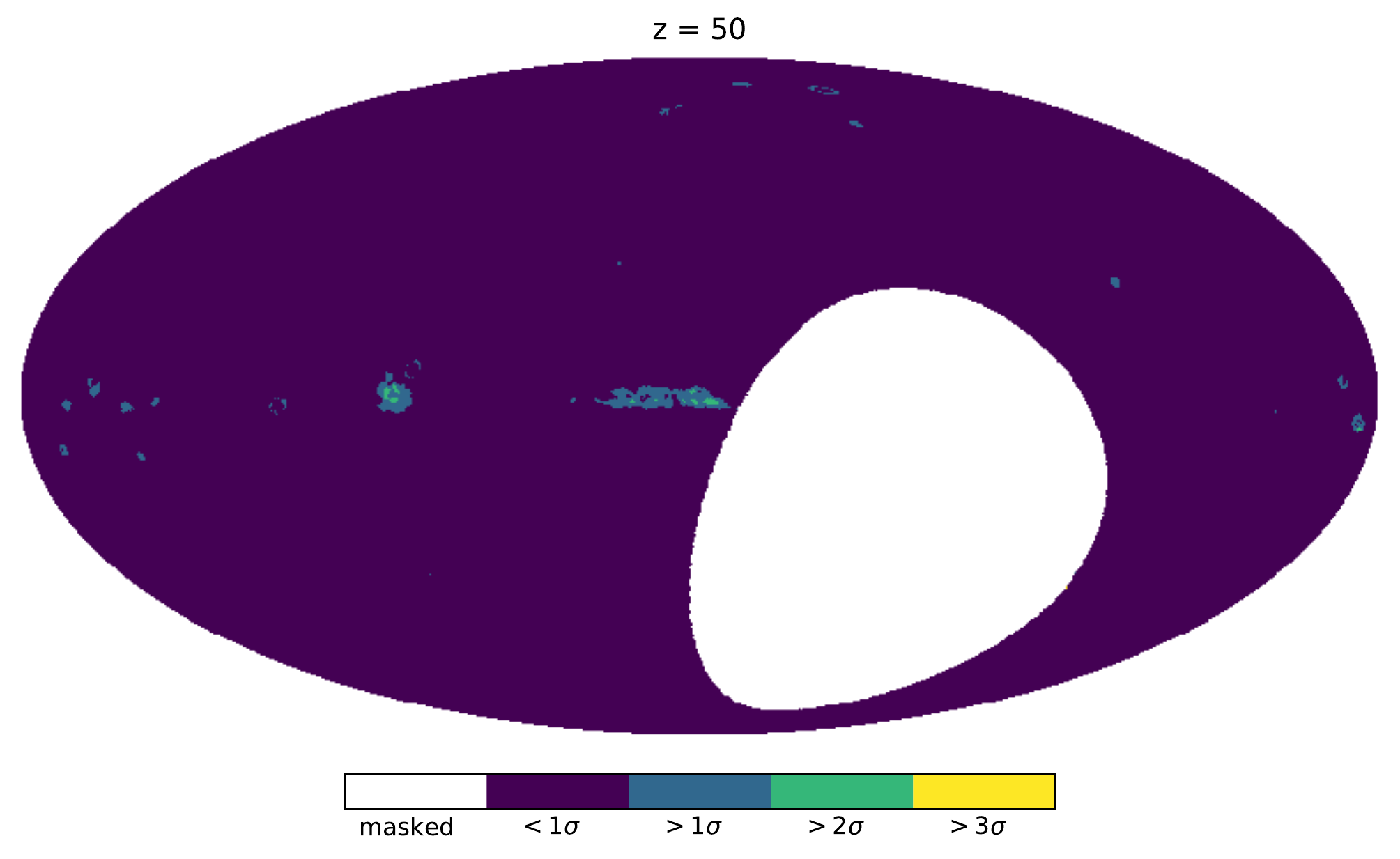}\label{fig:f3}
  \caption{Pixels that are optically thick to 21\,cm emission from redshifts $z=150$ (top), 100 (middle) and 50 (bottom).  Different colors represent the  statistical confidence with which we can claim the pixel is optically thick. As with Figure \ref{fig:fit_maps}, all maps are in Galactic coordinates.} 
  \label{fig:significance_maps}
\end{figure*}

\begin{table}[]
    \centering
    \begin{tabular}{c|c|c|c}
     \makecell{Percentage of the sky \\ optically thick at \dots} & $z=150$ & $z=100$ & $z=50$ \\
     \hline
        \dots > 1 $\sigma$ confidence & $27.7\%$ & $14.8\%$ & $0.719\%$ \\
        \hline
        \dots > 2 $\sigma$ confidence & $0.984\%$ & $0.697\%$ & $0.052\%$ \\
        \hline
        \dots > 3 $\sigma$ confidence & $0.189\%$ & $0.101\%$ & $0.001\%$ \\
      
    \end{tabular}
    \caption{: Percentage of the sky optically thick to the 21 cm signal with statistical confidence > $1\sigma$, $2\sigma$ and $3\sigma$ at redshift 150, 100 and 50.}
    \label{tab:my_label}
\end{table}

We also calculate what percentage of the whole sky is optically thick to 21 cm signal.  The results are presented in Table 1. Because a portion of the sky is masked (i.e. below the horizon for the LWA1), we first calculate the optical thickness ratio in Galactic and extragalactic regions. We calculated percentages of the sky which fall into the two regions in the unmasked and masked parts of the maps. $23\%$ of the masked region falls within 10 degrees of zero Galactic latitude, which we classified as Galactic region. However, only $16\%$ of the unmasked region falls into the Galactic region. To account for that, we calculated what percentage of  the unmasked region was optically thick in the Galactic region vs extragalactic region and assumed the same proportions in the masked region to calculate the final optically thick percentage of the whole  sky.

We then assumed the region of the Southern hemisphere covered by the horizon would have proportional optical thickness Galactic and extragalactic regions. This allowed us to estimate the optical thickness ratio in the whole sky.

\section{Discussion}
\label{sec:discussion}

The numbers in Table 1 are largely encouraging for high redshift 21\,cm cosmology, particularly for the lower redshifts of the cosmic Dark Ages. While challenges of foreground removal must still be overcome, the issue of foreground opacity appears of little concern. We do find that $\sim25 \%$ of the sky is optically thick for redshift $z=150$, albeit at low statistical significance. However, this does not appear to be random error.  Visual inspection of the data in Figure \ref{fig:pixelplot} shows that the LWA1 data do generally show a spectral flattening towards lower frequencies. Whether this effect is real or an artifact of the LWA1 sky survey is difficult to know.  We do note that the optical depth coefficient is always low in the regions of the sky corresponding to the horizons of the LWA1 site.  The response of the instrument is lower towards the horizon, making these higher noise regions, which we see in the LWA1 error maps.  However, the optical depth coefficient maps in the regions are not noisy but instead consistently lower than in the rest of the sky, suggesting there might be a primary beam correction issue that leads to altered frequency dependence at those low elevations. \citet{Dowell} also mentioned that LWA1 data has systemically lower temperatures near South Galactic Pole, which might be due to the limitations of in the dipole response model and the missing spacings corrections.  

Our results are generally consistent with older studies like \citet{Ellis&Hamilton}, which find that areas off the Galactic plane are optically thin at all but the lowest frequencies (most of which are not relevant for 21\,cm Cosmology), whereas in the plane of the Galaxy absorption can be significant at frequencies as high as 20\,MHz.  A more quantitative comparison with the sky model of \citet{Cong} reveals some interesting discrepancies, however.  We use their model to generate maps at 9.9 and 10.1\,MHz and look for pixels that (in specific intensity) are brighter at 10.1\,MHz than at 9.9\,MHz --- i.e., pixels that have a turnover in their spectrum due to free-free absorption at a frequency higher than 10\,MHz.  Depending on the specific model parameters used, we find that between 1 and 2\% of pixels meet this criterion.  Using our fits, however, we find that over 70\% of pixels have a spectral turnover at frequencies higher than 10\,MHz!  The reason for this seemingly significant discrepancy can be found by comparing the methodologies.  \citet{Cong} use low-frequency sky maps (including the LWA1 survey we analyze here) to calculate the spectral index of synchrotron emission in the limit of an \emph{absorption free} sky.  That is, they explicitly assume there is no effect from free-free absorption at any of the frequencies in the LWA1 sky maps, which they argue is justified based on the \texttt{NE2001} \citep{ne2001a,ne2001b} model of free electron density in the Galaxy.  In turn, any spectral turnovers in the \citet{Cong} model come from using the \texttt{NE2001} model to predict the optical depth along various lines of sight through the Galaxy.  Our methodology differs in that we fit a spectral model that includes a (potential) turnover directly to the LWA1 data.  As we have seen, many pixels in the LWA1 maps exhibit a spectral flattening towards the lowest frequencies, even off the plane of the Galaxy (see e.g. Pixels C and E in Figure \ref{fig:pixelplot}).  In effect, our fits interpret this spectral flattening as evidence of a significantly higher optical depth than is predicted by the \citet{Cong} model.

We again caution that this result is of low statistical significance given the relatively large uncertainties on the LWA1 data.  To estimate the statistical uncertainties on the above statement that 70\% of pixels in our fits have a spectral turnover at frequencies higher than 10\,MHz, we use the calculated error bars on the frequencies where the optical depth equals unity.\footnote{Strictly speaking, the frequency where the optical depth equals one is not the same as the frequency where the spectrum turns over (as can be seen in Figure \ref{fig:pixelplot}, the vertical dashed line does not always land at the peak of the best fit model spectrum).  However, calculating robust uncertainties on the turnover frequencies is more computationally demanding to Monte Carlo, since the exact position of the turnover depends on all three of our fit parameters (as opposed to just the optical depth coefficient, $F$, which alone determines the frequency at which the optical depth becomes unity).  As a check, we calculate both sets of uncertainties for the representative pixels in Figure \ref{fig:pixelplot} and find that the two generally agree.  Using one set of uncertainties as a proxy for the other should therefore be sufficient for the rough estimate presented here.}
When including these uncertainties, only $\sim$30\% of pixels have a spectral turnover above 10\,MHz with $1\sigma$ confidence; at $2\sigma$, this number drops to $\sim$1.5\%, consistent with the values calculated from the \citet{Cong} model.  Ultimately, we will need better and/or lower-frequency data to determine whether the spectral flattening seen at the lowest frequencies of the LWA1 survey is real and provides evidence of higher-than-expected free-free optical depths through the Galaxy.

If these low-significance results do hold, however, the impact on Dark Ages 21\,cm cosmology will be tangible --- but they certainly do not spell disaster for the field. For projects that envision the Dark Ages 21\,cm signal as the ultimate cosmological observable, where effectively all modes of the density power spectrum can be measured, this will reduce the total number of measurable modes and increase the effect of cosmic variance.  We do not provide detailed forecasting here, but even a $\sim25\%$ reduction in the number of measurements is unlikely to qualitatively change what can be done with Dark Ages observations given the tremendous potential of the technique.  (We note that CMB experiments like Planck recommend masking $\sim22\%$ of pixels, a very similar fraction of the sky; \citealt{planck}.)  The larger effect will likely be on the observational approach.  Experiments designed to map the 21\,cm signal can likely just mask optically thick regions from their analysis, akin to many CMB experiments.  
However, since the distribution of (potentially) optically thick regions is not random (c.f. Figure \ref{fig:significance_maps}), landing sites for lunar far side experiments should be carefully considering to minimize the fraction of observable sky that is obscured --- especially if, for simplicity of deployment, zenith-pointing, non-steerable antennas are used. Furthermore, experiments using non-imaging based techniques (e.g. \citealt{parsons_et_al_2012b}) will need to take care to exclude optically thick regions from their analysis, particularly if they have very large fields of view.

For additional data, we initially considered using OVRO-LWA (Owens Valley Long Wavelength Array) sky maps \cite{Eastwood_2018}, which were created using Tikhonov regularized $m$-mode analysis imaging techniques with angular resolution of $\sim15$ arcmin. However, these maps proved difficult to use in practice.  Unlike the LWA1 sky survey maps, no zero-level correction was performed.  We tried to calculate the offsets of each map ourselves by extrapolating the Haslam map at high Galactic latitudes to LWA frequencies to find offsets relative to each OVRO-LWA map. We also tried a similar approach using the 45\,MHz Guzman map \citep{Guzman}.  Ultimately, the results were overly sensitive on the offset correction, which emphasizes the importance of this procedure when using interferometric maps to make statements about spectral behavior of diffuse objects. Further complicating our use of the OVRO-LWA maps were the errors quoted as fixed amount of thermal noise (roughly $800\,\mathrm{mJy/beam}$ in each map) across the entire sky. Not only are these errors a factor of 10-20 less than the LWA1 errors, their uniformity across the sky led to unbelievable results of nominally high significance at low elevation angles. 

%We also considered using the existing global sky models for our analysis
%One of the leading model for the sky in the frequency range of 10\,MHz to 100\,GHz is the Global Sky Model \citep{GSM}. This model was extended to 30 MHz - 800 GHz range as eGSM (Kim & Liu, in prep.). However, as we're asking a very specific question about optical depths, it was better served using raw data. Global sky models are more general and we would have to find a way to deal with correlations vs frequency and the effects of other assumptions that go into making a Global sky model.

Lower frequency data can answer these questions, but because of the ionosphere it becomes more difficult to get high-quality maps below 30\,MHz.  Lower frequency observations from space and/or the lunar far side may provide the best foreground maps for this kind of analysis. 
Another issue seen in our analysis (not directly related to the data quality) is that the fit seems to break down in the very center of the Galactic plane. This does not come as a surprise, as the two-component model is appropriate for much of the sky ``except near the  galactic plane where higher intensities are encountered'' \citep{TMS}. As these are H-II regions, dominated by bremsstrahlung radiation, our two-component model will not be able to describe it accurately.  However, we are not worried about falsely interpreting these pixels because we can be quite confident that the center of the Galaxy is optically thick.
In addition, the optical depth coefficient values in this pixels don't have a significant effect on the overall results as they comprise only a small fraction of the sky.
We also note that we did not consider extragalactic free-free absorption. Given the limitations of the LWA1 data, we do not expect that we could distinguish this contribution to any spectral turnover from absorption with a Galactic origin even if it was included as another free parameter in our fits. However, higher-precision future studies with lower-frequency data could be used to explore the impact of this effect. 
%We are limited by the data we have but other possible data sources also have their issues.

\section{Conclusion}
\label{sec:conclusion}

We used the LWA1 \citep{Dowell} dataset at 9 frequencies combined with the 408\,MHz Haslam map \citep{Haslam_1981, Haslam_1982, Remazeilles} to fit a two-component model from \citet{Cane} and calculate the optical depth of foregrounds at low frequencies. We used a nonlinear least squares fit to find 3 parameters of that model and calculated the turnover frequency of each pixel. We then used Monte Carlo simulation to calculate the error  bars of  the turnover frequency to find which regions of the sky are optically thick  to 21\,cm signal at redshifts 150, 100 and 50, respectively. In conclusion, our results are very encouraging for Dark Ages 21\,cm cosmology as most of the sky is not optically thick at high statistical significance. However, at the highest redshifts ($z=150$), we see low significance evidence that a large portion of sky may be optically thick to the 21\,cm signal. Lower-frequency maps are likely the best way to resolve this issue.

\section*{Acknowledgements}

The authors would like to thank Adrian Liu for several exceptionally helpful conversations related to this project, Avery Kim for providing a HEALPix version of the Guzman 45\,MHz map, and Yanping Cong for help installing their ULSA code.  We would also like to thank the LWA1 Low Frequency Sky Survey team for publicly releasing their maps, and our anonymous referee for helping to improve the manuscript.  This research was conducted using computational resources and services at the Center for Computation and Visualization, Brown University.

%%%%%%%%%%%%%%%%%%%%%%%%%%%%%%%%%%%%%%%%%%%%%%%%%%
\section*{Data Availability}

The data underlying this article were accessed from LWA1 Low Frequency Sky Survey: \url{http://lda10g.alliance.unm.edu/LWA1LowFrequencySkySurvey/} and the desourced and destriped 408\,MHz Haslam map: \url{https://lambda.gsfc.nasa.gov/product/foreground/fg_2014_haslam_408_info.cfm} The derived data generated in this research will be shared on reasonable request to the corresponding author.



\bibliographystyle{mnras}
\bibliography{bibliography} % if your bibtex file is called example.bib

%\begin{thebibliography}{99}
%\end{thebibliography}

%%%%%%%%%%%%%%%%%%%%%%%%%%%%%%%%%%%%%%%%%%%%%%%%%%

%%%%%%%%%%%%%%%%% APPENDICES %%%%%%%%%%%%%%%%%%%%%

\appendix

%%%%%%%%%%%%%%%%%%%%%%%%%%%%%%%%%%%%%%%%%%%%%%%%%%

% Don't change these lines
\bsp	% typesetting comment
\label{lastpage}
\end{document}